# Asynchronous Early Output Block Carry Lookahead Adder with Improved Quality of Results

P. Balasubramanian, D.L. Maskell
School of Computer Science and Engineering
Nanyang Technological University
Singapore 639798

N.E. Mastorakis
Department of Industrial Engineering
Technical University of Sofia
Sofia 1000, Bulgaria

*Abstract*—A new asynchronous early output block carry lookahead adder (BCLA) incorporating redundant carries is proposed. Compared to the best of existing semi-custom asynchronous carry lookahead adders (CLAs) employing delay-insensitive data encoding and following a 4-phase handshaking, the proposed BCLA with redundant carries achieves 13% reduction in forward latency and 14.8% reduction in cycle time compared to the best of the existing CLAs featuring redundant carries with no area or power penalty. A hybrid variant involving a ripple carry adder (RCA) in the least significant stages i.e. BCLA-RCA is also considered that achieves a further 4% reduction in the forward latency and a 2.4% reduction in the cycle time compared to the proposed BCLA featuring redundant carries without area or power penalties.

## I. INTRODUCTION

The International Technology Roadmap on Semiconductors [1] highlights variability as one of the grand challenges for electronic design in the nanoelectronics regime. To cope with the crucial design issues such as modularity, parameters uncertainty, variability etc. the asynchronous design method is a practical alternative to the synchronous design method. Especially, asynchronous circuits employing delay-insensitive data encoding and adhering to a 4-phase handshake protocol are generally robust as they can innately absorb variations in process, voltage, and temperature parameters etc. [2]. Such asynchronous circuits are usually elastic and are commonly referred to as quasi-delay-insensitive (QDI). QDI circuits are the practically implementable delay-insensitive asynchronous circuits with the only exception and assumption of isochronic forks [3]. An isochronic fork refers to the simultaneous acknowledgment of a signal transition on all the forks arising from a node or junction. Isochronic forks represent the weakest compromise to delay-insensitivity.

Given that an adder is an important datapath element of any general purpose or digital signal processing unit, the efficient design of an asynchronous adder is of interest and importance. References [4 – 7] present different QDI designs of the high-speed carry lookahead adder (CLA). All these are ASIC-based designs. Reference [4] adopts the full-custom design approach using transistors while [5 – 7] adopt a logic synthesis driven semi-custom design approach by utilizing the gates of a standard cell library and including the C-element (which is depicted using the circle with the marking C in the Figures). The C-element[1] is widely used in asynchronous circuit designs but does not form a part of a commercial standard cell library. Hence the C-element should be custom-designed for use in asynchronous logic synthesis. Different realizations of the Muller C-element are given in [8].

## II. ASYNCHRONOUS CIRCUIT DESIGN – FUNDAMENTALS

The principles of asynchronous design are discussed here.

### A. Delay-Insensitive Dual-Rail Data Encoding and 4-Phase RTZ Handshaking

As per the dual-rail (i.e. 1-of-2) code, a single-rail binary input say W, is encoded using two wires as say, W1 and W0. The valid data is represented as follows: W = 1 is represented by W1 = 1 and W0 = 0. W = 0 is represented by W1 = 0 and W0 = 1. W1 and W0 cannot assume 1 concurrently as it is illegal and invalid since the coding scheme would become unordered. However, W1 and W0 can assume 0 concurrently, and this assignment is called the spacer.

An asynchronous circuit stage that employs the delay-insensitive dual-rail code for data encoding and processing and the 4-phase return-to-zero (RTZ) handshake protocol for data communication is shown in Fig. 1. As the name implies, the 4-phase RTZ handshake protocol consists of four phases which will be explained with reference to Fig. 1 by assuming dual-rail encoded data although this explanation would be applicable for data represented using any delay-insensitive 1-of-*n* code [2].

In the first phase, the dual-rail data bus shown in Fig. 1, specified by (W1, W0) etc. is in the spacer state and ACKIN is 1. The transmitter transmits a code word (a valid data) and this results in rising signal transitions i.e. binary 0 to 1 on one of the corresponding dual rails of the dual-rail data bus. In the second phase, the receiver receives the code word sent, and it drives ACKOUT to 1. In the third phase, the transmitter waits for ACKIN to become 0 and then resets the entire dual-rail data bus to the spacer state. In the fourth phase, after an unbounded positive and finite time duration, the receiver drives ACKOUT to 0 i.e. ACKIN becomes 1. One data transaction is now said to be complete, and the asynchronous circuit stage can commence the next data transaction. Thus, the application sequence of the input data is: valid data-spacer-valid data-spacer, and so forth.

---
[1] The C-element outputs 1 or 0 if all its inputs are 1 or 0 respectively. It will maintain its existing steady-state if the inputs are not the same.

This work is supported by the Singapore Ministry of Education (MoE) Academic Research Fund Tier 2 under grant MOE2017-T2-1-002 and MoE Tier 1 under grant RG132/16.

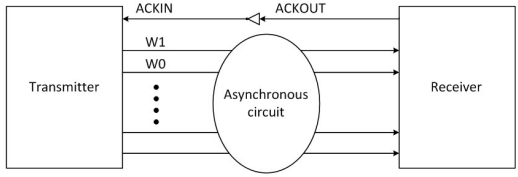

Fig. 1. An asynchronous circuit stage

*B. Asynchronous Circuit Types*

Asynchronous circuits are generally categorized as strongly indicating [9], weakly indicating [9], and early output [10]. Indication implies providing acknowledgment for the receipt of the primary inputs through the primary outputs. Indication should also be provided by the intermediate circuit outputs [2]. The timing characteristic of strong indication, weak indication, and early output asynchronous circuits is illustrated in Fig. 2.

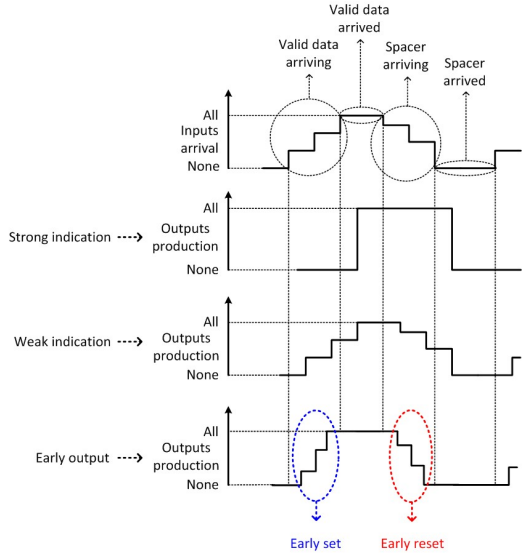

Fig. 2. Input-output timing characteristics of various asynchronous circuits

A strong indication asynchronous circuit would start data processing to produce the required primary outputs only after receiving all the primary inputs whether they are valid data or spacer. A weak indication asynchronous circuit would start data processing and could produce some of the primary outputs after receiving a subset of the primary inputs. Nonetheless, the production of at least one primary output is delayed till the last primary input is received. An early output asynchronous circuit could start data processing and produce all the primary outputs after receiving a subset of the primary inputs. If all the valid (spacer) primary outputs are produced after receiving valid data (spacer) on a subset of the primary inputs, the early output asynchronous circuit is said to be of early set (early reset) type. The early set and reset behaviors are depicted within the blue and red ovals in dotted lines in Fig. 2. Relative-timed circuits [11] form a subset of the early output circuits but incorporate extra timing assumptions to guarantee their safe operation.

Strong indication arithmetic circuits would encounter the worst-case latency for processing valid data and spacer. But weak indication, early output and relative-timed arithmetic circuits would encounter data-dependent latency for processing the valid data and may experience only a constant latency for processing the spacer. Therefore, the latter categories are rather preferable to physically implement computer arithmetic [19].

III. PROPOSED EARLY OUTPUT BCLA AND BCLA-RCA

An early output BCLA is constructed using three building blocks namely the block carry lookahead generator (BCLG), the full adder (FA), and the sum logic (SL). The proposed 4-bit early output BCLG incorporating regular and redundant carry outputs, the early output FA [10], and the early output SL [10] are shown in Figs. 3a, 3b and 3c respectively. In Fig. 3a, (A01, A00) to (A31, A30), and (B01, B00) to (B31, B30) represent the dual-rail encoded augend and addend inputs while (C01, C00) represents the dual-rail encoded carry input. The regular and redundant dual-rail carry outputs are specified by (C41, C40), and (RedC41, RedC40) respectively.

The equations for the 4-bit BCLG regular carry outputs are expressed by (1) and (2). The logic equations for the redundant carry outputs are the same as (1) and (2) although the logic realization of (RedC41, RedC40) is different from (C41, C40) in the last logic level. In (1) and (2), P3 to P0 represent the carry-propagate signals, G3 to G0 represent the carry-generate signals, and K3 to K0 represent the carry-kill signals. Their equations may be interpreted from Fig. 3a. Note that (1) and (2) are in the disjoint sum-of-products (DSOP) form [12] [13]. In a DSOP expression, the products are all mutually orthogonal and the logical conjunction of any two product terms yields 0.

$$C41 = G3 + P3G2 + P3P2G1 + P3P2P1G0 + P3P2P1P0C01 \quad (1)$$

$$C40 = K3 + P3K2 + P3P2K1 + P3P2P1K0 + P3P2P1P0C00 \quad (2)$$

In Fig. 3b, (A1, A0), (B1, B0) and (CIN1, CIN0) represent the dual-rail encoded augend, addend and carry inputs of the FA, and (SUM1, SUM0) and (COUT1, COUT0) represent the respective dual-rail encoded sum and carry outputs. The logic realization of SL, shown in Fig. 3c, is identical to the FA except for the dual-rail encoded carry output.

The respective architectures of 32-bit early output BCLAs with regular carry outputs, and regular and redundant carry outputs are shown in Figs. 4a and 4b. Figs. 4c and 4d show the hybrid BCLA-RCA architectures with only regular, and regular and redundant carry outputs. In a BCLA [14], a block refers to a sub-section of the CLA. Unlike a recursive CLA (RCLA) [6] where carries between blocks are rippled and carries within blocks are generated by lookahead, in a BCLA carries ripple within blocks and carries between blocks are generated by lookahead. In this context, the BCLA can also be termed as the section-carry based CLA i.e. SCBCLA [5]. In an *m*-bit sub-RCLA, *m* lookahead carry outputs are produced whereas in an *m*-bit sub-BCLA or sub-SCBCLA, only one lookahead carry output for the successive stage sub-BCLA/sub-SCBCLA is produced. The hybrid BCLA-RCA architecture utilizes an RCA in the least significant stages to compensate for the maximum propagation delay encountered in the least significant BCLG, and this may reduce the datapath delay.

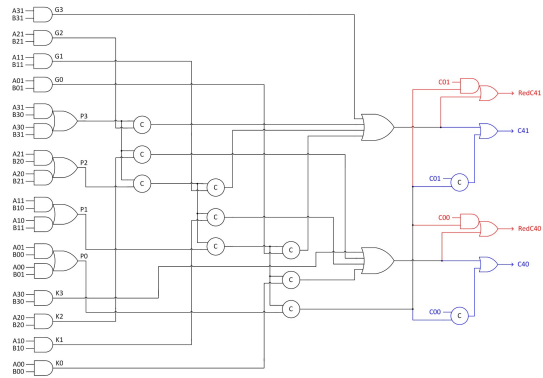

**Fig. 3a. 4-bit early output BCLG with regular and redundant carry outputs**

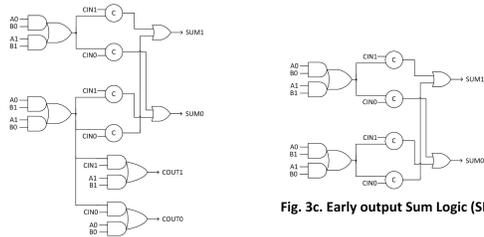

**Fig. 3b. Early output Full Adder (FA)**

**Fig. 3c. Early output Sum Logic (SL)**

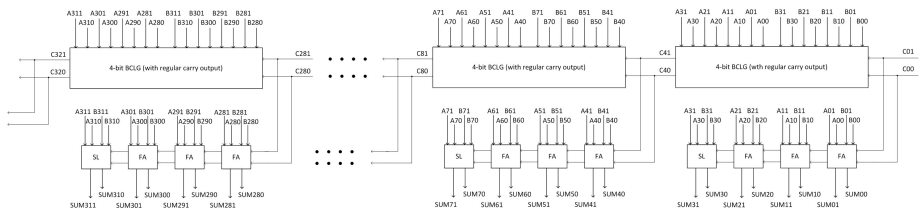

**Fig. 4a. 32-bit early output BCLA with regular carry outputs**

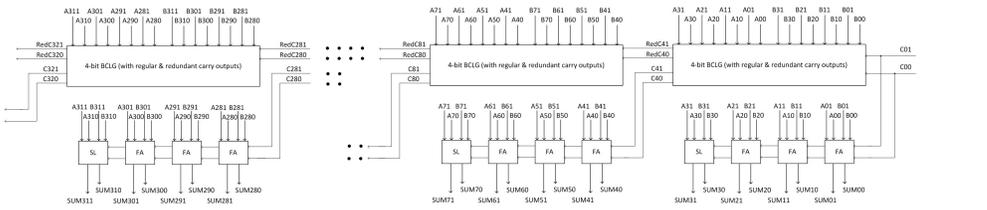

**Fig. 4b. 32-bit early output (i.e. relative-timed) BCLA with regular and redundant carry outputs**

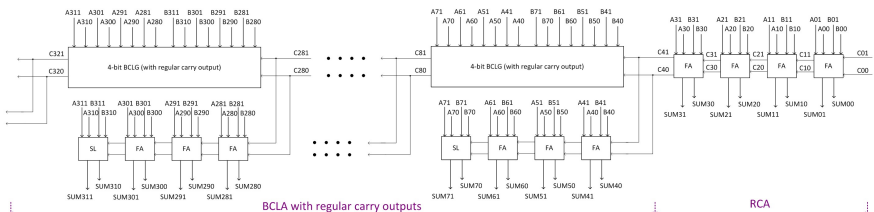

**Fig. 4c. 32-bit early output BCLA (with regular carry outputs) and RCA hybrid**

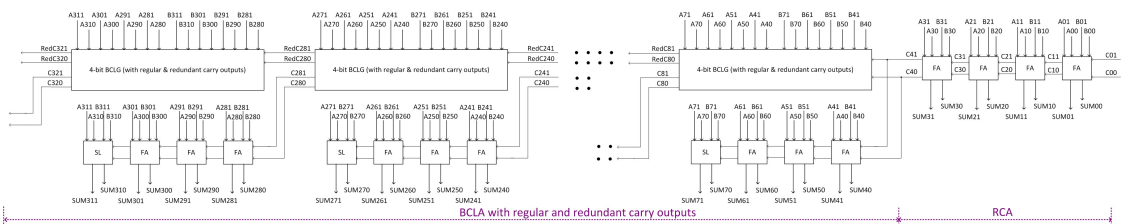

**Fig. 4d. 32-bit early output (i.e. relative-timed) BCLA (with regular and redundant carry outputs) and RCA hybrid**

In Figs. 4b and 4d, early output asynchronous sub-BCLAs with regular and redundant carry outputs are shown. Redundant carries propagate between successive BCLGs while regular carries ripple and enable to produce the sum bits. The final level of logic that generates the redundant carry outputs, shown in red in Fig. 3a, is better optimized for speed compared to the final level of logic that generates the regular carry outputs shown in blue in Fig. 3a. Forward latency in an asynchronous circuit refers to the time taken for processing the valid data, and the reverse latency refers to the time required for processing the spacer. The 'cycle time' is the sum of forward and reverse latencies and is an important metric as it governs the rate at which fresh data can be input to an asynchronous circuit. Redundant carry outputs serve two important purposes of reducing the reverse latency and the cycle time significantly [15]. This is the reason for the reduced reverse latency encountered for the application of the spacer in Figs. 4b and 4d. Forward and reverse latencies of BCLA (Fig. 4a) and hybrid BCLA-RCA (Fig. 4c) with regular carries are data-dependent. On the other hand, the forward latencies of BCLA (Fig. 4b) and hybrid BCLA-RCA (Fig. 4d) with regular and redundant carries are data-dependent but their reverse latencies are just a constant.

## IV. RESULTS AND CONCLUSION

Many 32-bit asynchronous CLAs and hybrid CLA-RCAs described in the literature were physically implemented based on a 32/28nm CMOS digital cell library [16]. Input and output registers, and the completion detector are kept identical for all the CLAs and CLA-RCAs. Hence the diverse CLAs and CLA-RCAs differ only in their logic. Therefore, the differences between their design metrics can be directly attributed to the differences between their logic compositions. About 1000 random input vectors were identically supplied to all the adders to verify their functionalities and to capture their respective switching activities. Then the average power was estimated. Area, forward latency, cycle time, and average power dissipation were estimated for the CLAs using Synopsys tools and are given in Table I. Forward latency is akin to critical path delay and is directly estimated. The reverse latency was estimated based on the gate-level simulation timing data.

The results are split into five groups, as shown in Table I. Group1 results were obtained by utilizing the early output 4-bit SCBCLG of [5] and the weak indication FA and SL of [17]. Group2 results were obtained by utilizing the early output 4-bit SCBCLG of [5] and the weak indication FA and SL of [18]. Group3 results were obtained by using the early output 4-bit sub-RCLA of [6] and the early output FA and SL of [10]. Group4 results were obtained by utilizing the early output 4-bit SCBCLG of [7] and the early output FA and SL of [10]. Group5 results were obtained using the proposed 4-bit BCLG and the early output FA and SL of [10].

From Table I, it is seen the proposed BCLA featuring regular and redundant carries achieves less forward latency and cycle time compared to the other CLAs without or with the redundant carries besides the regular carries. Compared to the Group4 SCBCLA with redundant carries, which reports the least cycle time among the existing designs reported in Table I, the proposed BCLA with redundant carries achieves 13% reduction in forward latency and 14.8% reduction in cycle time with no area or power penalty. The proposed hybrid BCLA-RCA featuring regular and redundant carries achieves further reductions in the forward latency and cycle compared to the proposed BCLA featuring redundant carries by 4% and 2.4% respectively, again, with no penalty in terms of area or average power dissipation.

TABLE I. DESIGN METRICS OF ASYNCHRONOUS CLAs/CLA-RCAs

| Results Group | Adder Type | Latency[†] (ns) | Cycle Time (ns) | Area ($\mu m^2$) | Power ($\mu W$) |
|---|---|---|---|---|---|
| Group1 [5, 17] | SCBCLA[ξ] | 3.31 | 6.24 | 2951.88 | 2191 |
| | SCBCLA* | 2.46 | 4.15 | 2987.46 | 2192 |
| Group2 [5, 18] | SCBCLA[ξ] | 3.14 | 6.02 | 2915.29 | 2188 |
| | SCBCLA* | 2.32 | 4.00 | 2950.87 | 2189 |
| Group3 [6, 10] | RCLA[ξ] | 2.75 | 5.50 | 2569.65 | 2177 |
| Group4 [7, 10] | SCBCLA[ξ] | 3.13 | 6.01 | 2524.92 | 2178 |
| | SCBCLA* | 2.31 | 3.98 | 2560.50 | 2179 |
| Group5 (This Work) | BCLA[ξ] | 2.76 | 5.26 | 2209.78 | 2174 |
| | BCLA* | 2.01 | 3.39 | 2245.36 | 2176 |
| | BCLA-RCA* | 1.93 | 3.31 | 2171.41 | 2174 |

[†] Forward latency; [ξ] Regular carries; * Regular and Redundant carries